\documentclass[preprint,12pt]{elsarticle}

\usepackage{amssymb}

\usepackage{amsthm}

\journal{Scripta Materialia}

\begin{document}

\begin{frontmatter}



\title{Localised stress and strain distribution in sliding}


\author[a]{Anna Kareer }
\author[c]{Eralp Demir}
\author[c]{Edmund Tarleton}
\author[b]{Christopher Hardie}

\affiliation[a]{organization={Department of Materials, University of Oxford},
           addressline={Parks Road},
             city={Oxford},
            postcode={OX1 3PH},
             country={United Kingdom}}

\affiliation[b]{organization={Culham Centre for Fusion Energy, UK Atomic Energy Authority},
             city={Abingdon},
            postcode={OX14 3DB},
             country={United Kingdom}}
\affiliation[c]{organization={Department of Engineering Science, University of Oxford},
           addressline={Parks road},
             city={Oxford},
            postcode={OX1 3PJ},
             country={United Kingdom}}

\begin{abstract}
In this paper, we present a comprehensive analysis of the contact mechanics associated with a micron-sized sliding asperity, which plays a crucial role in the abrasive wear processes. Utilising nanoscratch testing, we experimentally investigate the deformation and employ High-Resolution Electron Backscatter Diffraction (HR-EBSD) to characterise the resulting strain fields at various locations in the residual nanoscratch. To simulate these experiments, we utilise a physically-based Crystal Plasticity Finite Element (CPFE) model, enabling a three-dimensional simulation that can accurately capture the measured elastic and plastic strain fields around the sliding contact. This knowledge serves as a foundation from which we may be able to discern the multi-physical processes governing micro-scale wear phenomena. 
\end{abstract}



\end{frontmatter}



Despite centuries of research and advancements in the field of tribology, friction and wear still cause substantial economic losses in terms of both energy and resources \cite{jost1976economic, holmberg2017influence}. While friction is commonly perceived to incur higher direct costs than wear, it has been found that maintenance expenses from wear can account for up to 15\% of all costs, with a considerable portion attributed to replacing worn components \cite{holmberg2017influence}. Effectively controlling and predicting the wear behaviour of tribosystems, holds the potential for developing materials with enhanced wear resistance, yielding substantial economic benefit and environmental sustainability across a range of engineering sectors. 

The complexity of wear prediction arises due to the range of multi-physical wear mechanisms that take place at the interface between contacting surfaces \cite{lim1987wear,lim1987overview}. It is well accepted that wear is governed by the interaction between micron sized surface asperities that come into contact and mechanically deform the counter surface and that frictional energy dissipation primarily results from the plastic deformation of this near-surface layer from contact between asperities \cite{bowden2001friction}. Therefore, a more appropriate predictive tool would be based on the microscale asperity interactions and allow discrete physical mechanisms to be selected and integrated, so that the wear behaviour, for various materials and operating conditions, may be accurately quantified. A model as such must be based on fundamental principles and we must first gain a mechanistic understanding of microscale sliding.     

Plastic deformation induced by the external tribological load that couples the normal and tangential load is responsible for a range of phenomena observed in microscale wear. Subsurface plastic deformation causes microstructural evolution that leads to a layer of material at the surface that differs to the bulk material \cite{rigney1979plastic}. By considering the single asperity problem, Greiner et al. aimed to identify the source of this surface discontinuity. They hypothesised that the inhomogeneous stress field associated with sliding, causes dislocations to self organise in a line below and parallel to the surface \cite{Greiner2018TheOO, greiner2016sequence, Xu2021OnTO}. The tribological load also dictates whether a material will deform by plasticity or by partial slip of the interface, a mechanism that has been attributed to the `stick-slip' phenomena commonly associated with frictional contacts. Partial slip arises due to stress singularities at the edge of the sliding contact that are not observed for normal loading alone \cite{brazil2021direct}. High-Resolution Electron Backscatter Diffraction (HR-EBSD) combined with Crystal Plasticity Finite Element Methods (CPFEM) has been used to study the lattice rotation fields under a sliding contact, generated using nanoscratch experiments, in a single crystal of copper \cite{kareer2020scratching}. The rotation fields display an inner and an outer rotation zone of opposing direction; the inner rotation zone is formed during the indentation process, and the outer develops as a result of the frictional contact giving some indication as to why the nanoscratch hardness property is significantly higher than that in from indentation \cite{kareer2016existence, kareer2016interaction}. In any attempt to interpret and predict these microscale wear phenomena, we first must fully understand the imposed stress distribution and strain generated during sliding. Apart from analytical models of the elastic contact, the stress distribution imposed on the surface by a tribological load is unkown \cite{hamilton1983explicit}. In this work we report the residual strain and stress fields generated around a sliding contact. Nanoscratch experiments are used to emulate the single asperity problem, and a crystal plasticity finite element (CPFE) model is used to simulate the deformation fields, in three-dimensions. This work provides a comprehensive description of deformation generated by a tribological load, which is of both fundamental scientific and practical importance. 

Nanoscratch experiments using a Berkovich tip in the edge forward orientation, were made into an $(001)$ surface of single crystal copper. The scratch was generated by simultaneously indenting the surface using an applied normal force of 3 $mN$, and laterally moving the stage and sample (in the x-direction) to create a 100 $\mu m$ long scratch. An initial indentation was made in the surface, and once the maximum load was reached, the sample stage was traversed laterally, creating a scratch. The tangential force and normal displacement were measured during scratching. Once the scratch had traversed 100 $\mu m$, the normal load was removed from the stationary stage. The single crystal was oriented such that the direction of motion during sliding was parallel to the $[100]$ crystallographic direction. $3 \mu m$ thick lamella were prepared and extracted using a focused ion beam (FIB), to study the subsurface deformation. EBSD maps were collected from the deformed surface with $(001)$ plane normal, a cross-section perpendicular to the scratch direction with $(\overline{1}00)$ plane normal and a cross-section parallel to the scratch direction with $(0\overline{1}0)$ plane normal. HR-EBSD analysis was used to capture the experimental, residual elastic strain and stress distribution in the planes of the cross-sections and at the surface \cite{wilkinson2006high, wilkinson2010determination, britton2012high}. Finite element simulations were performed using Abaqus 2020 to investigate the three-dimensional mechanics of deformation. A crystal plasticity user material (UMAT) for Abaqus was used based on the crystal plasticity solver scheme originally proposed by Dunne et al. \cite{dunne2007lengthscale, dunne2012crystal}. The constitutive laws in the model were calibrated against the experiments. Full details of the nanoscratch experiments, cross-section preparation, EBSD/HR-EBSD analysis, and the constitutive laws and parameters used in the CPFE model are provided in \cite{kareer2020scratching} where we reported the lattice rotation fields from these experiments.  

\begin{figure}[hbt!]
\centering
\includegraphics[width=\textwidth]{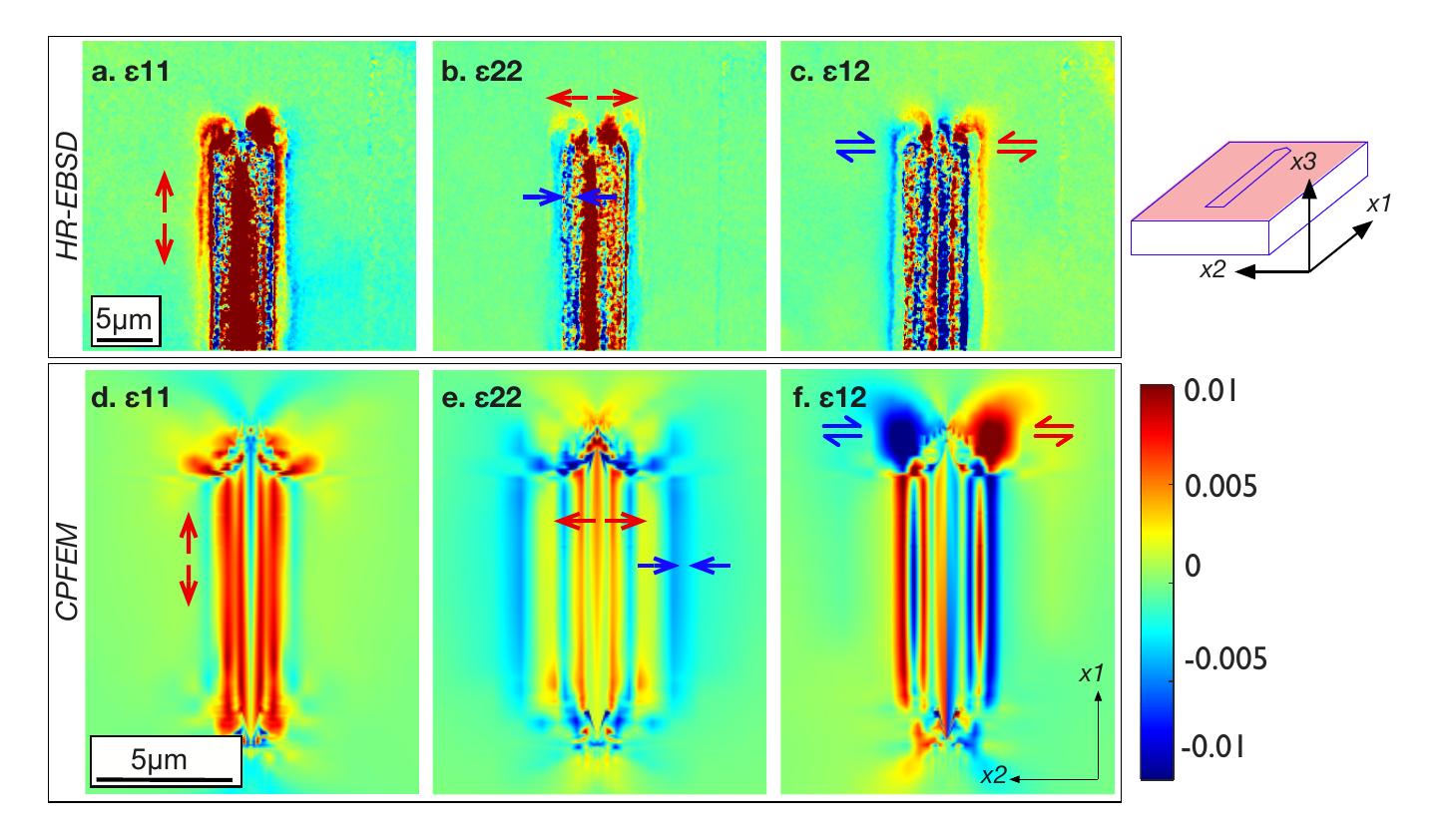}
\caption{Elastic strain fields at the surface with (100) plane normal measured by HR-EBSD (a., b. and c.) and simulated using CPFEM (d., e. and f.). Colour code shows the elastic strain around the scratch. The schematic in the top right shows the co-ordinate system used, where positive $x_1$ is the scratch direction.}
\label{straintop}
\end{figure}

\begin{figure}[hbt!]
\centering
\includegraphics[width=\textwidth]{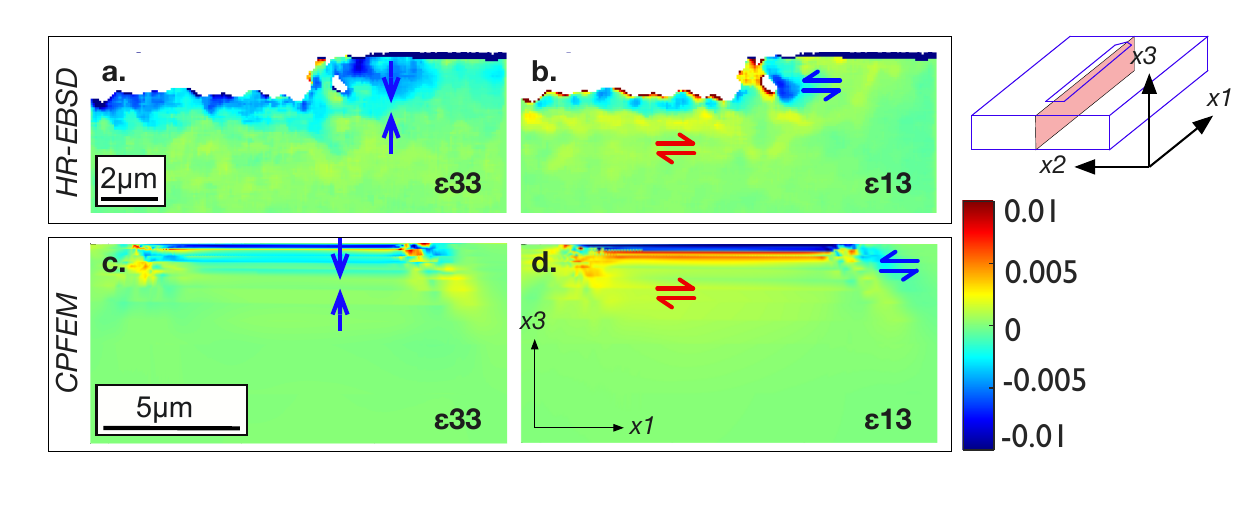}
\caption{Elastic strain fields in a cross-section plane parallel to the scratch direction with $(0\overline{1}0)$ plane normal measured by HR-EBSD (a., b.) and simulated using CPFEM (c., d.). Colour code shows the elastic strain around the scratch. The schematic in the top right shows the co-ordinate system used, where positive $x_1$ is the scratch direction and the shaded area is the cross-section plane.}
\label{strain slice}
\end{figure}

\begin{figure}[hbt!]
\centering
\includegraphics[width=\textwidth]{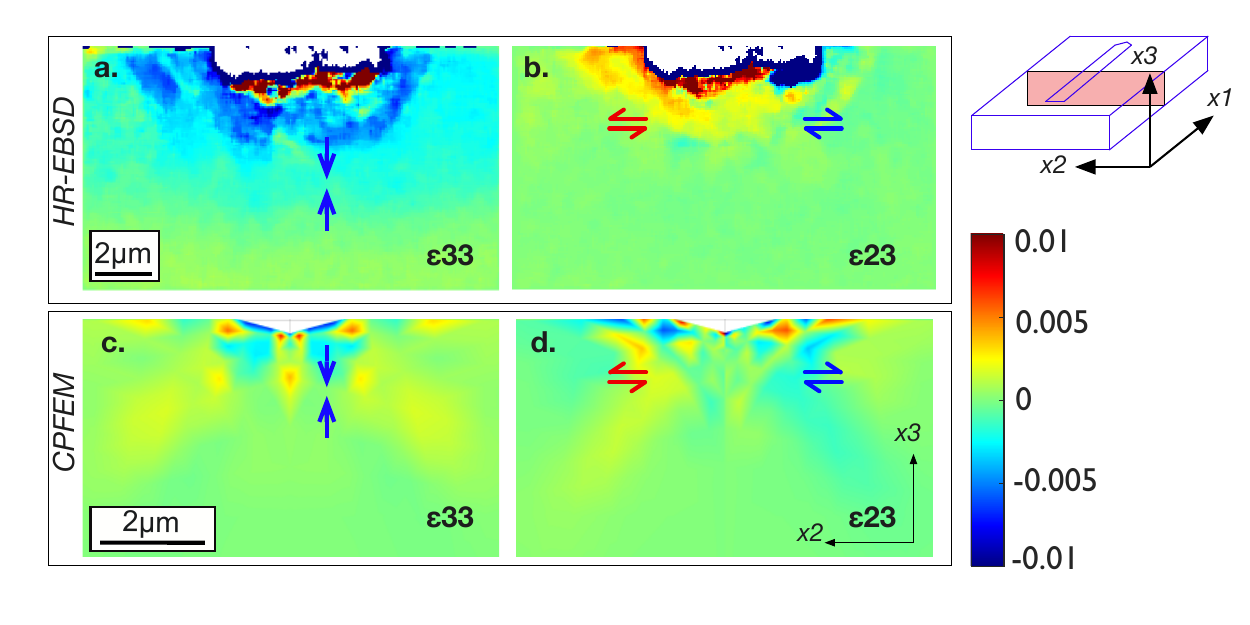}
\caption{Elastic strain fields in a cross-sectional plane perpendicular to the scratch direction with $(\overline{1}00)$ plane normal measured by HR-EBSD (a., b.) and simulated using CPFEM (c., d.). Colour code shows the elastic strain around the scratch. The schematic in the top right shows the co-ordinate system used, where positive $x_1$ is the scratch direction and the shaded area is the cross-section plane.}
\label{strain xs}
\end{figure}

Figures \ref{straintop} - \ref{strain xs} show the elastic strain fields measured experimentally and simulated by the CPFE model for the surface and cross-sectional planes. Note that the regions of severe plastic deformation, i.e. close to the indenter apex, in the subsurface cross-sections have not been included in this data due to a high distortion of the diffraction patterns within this region, which results in a poor quality cross-correlation with the reference pattern. In all cases the sign and magnitude of the simulated strain fields correlate well with those measured experimentally.   

Figure \ref{straintop} shows the elastic strain fields at the surface. Data taken from the surface shows an elastic tensile strain in direction of scratching (Figure \ref{straintop} a. and d.) in the direction perpendicular to the scratch, a tensile strain is observed inside the scratch track, with regions of compressive strain to support the residual piled-up material at the edges of the scratch (Figure \ref{straintop} b. and e.). The large in-plane elastic shear strains of alternating sign (Figure \ref{straintop} c. and f.) are formed due to the plastic deformation and pile-up of material both at the edges and ahead of the scratch. Two large regions of shear strains exist ahead of the leading facets of the tip where the final point of contact was made in the scratch, before the indenter was unloaded. Figure \ref{strain slice} and Figure \ref{strain xs} show the residual elastic strains in the cross-section with $(0\overline{1}0)$ and $(\overline{1}00)$ plane normal respectively. Subsurface, a residual compressive strain exists in the direction of loading (a. and c. in Figures \ref{strain slice} and \ref{strain xs}). The in-plane shear strains show two bands of shear strain with opposite sign directly beneath and parallel to the scratch (Figure \ref{strain slice} b., d.). The simulation captured a large region of compressive strain with a distribution of localised points of tensile strain (Figure \ref{strain xs} c.). The in-plane shear strains measured with HR-EBSD (Figure \ref{strain xs} b.) show the far field shear strains that are captured by the simulation (Figure \ref{strain xs} d.). In the model however, it can be seen that at the edge of the scratch track, regions of opposing shear strains are formed to support the redistributed material.  

The elastic strain fields presented above, combined with the lattice rotations fields presented in \cite{kareer2020scratching}, allow a full interpretation of the residual deformation from a microscale sliding contact. The deformation in sliding appears to extend further than that in an indentation and  displays a more complex pattern of opposing shear strains and lattice rotations. 

\begin{figure}[hbt!]
\centering
\includegraphics[width=0.7\textwidth]{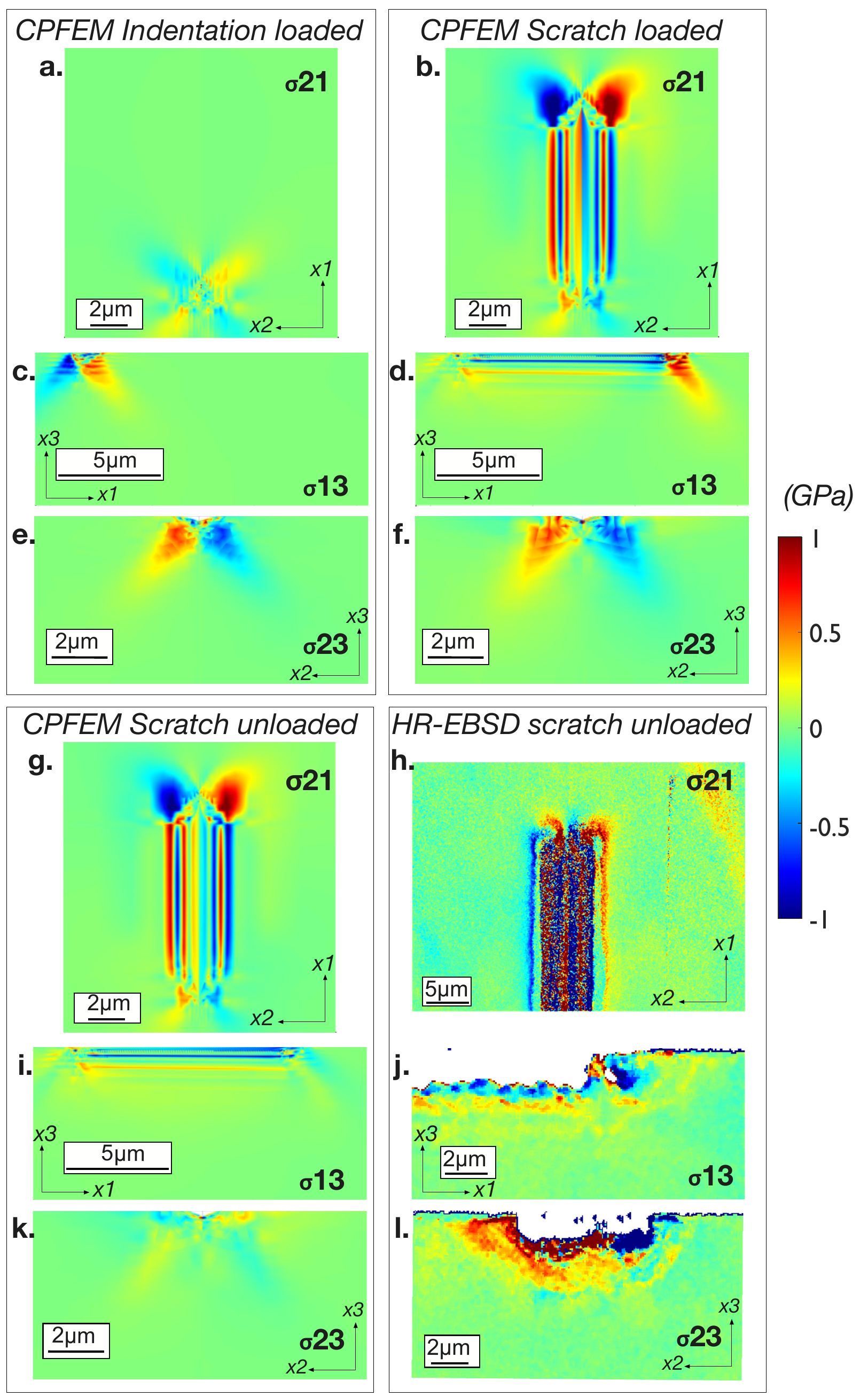}
\caption{Simulated shear stress distribution around a scratch for the initial indentation (a., c., e.) the full scratch before unloading (b., d., f.) and the unloaded scratch (g., i., k.). The stress distribution measured using HR-EBSD is given for the unloaded scratch is given in h., j. and l.}
\label{stress}
\end{figure}

To gain a mechanistic understanding of this deformation from sliding, we consider the simulated stress distribution during contact. Figure \ref{stress} shows the simulated stress distribution for an indentation under load (Figure \ref{stress} a., c., e.), the completed scratch under load (Figure \ref{stress} b., d., f.) and the unloaded scratch (Figure \ref{stress} g., i., k.). The unloaded, experimentally measured stress fields from the HR-EBSD are presented and show good agreement with the simulation (Figure \ref{stress} h., j., l.). For the indentation, as expected, large zones of shear stresses are formed at the facets of the indenter where they make contact with, and deform the surface. Subsurface, these shear stresses are equal in magnitude either side of the indenter apex following the symmetry of the contact. Note that due to the geometry of the indenter with respect to the $x_{13}$ plane, there is a slightly larger region of shear stress observed to the right of the apex (positive shear in (Figure \ref{stress} c.). 

As the scratch progresses, larger regions of shear stress are observed subsurface in the loaded scratch compared to the loaded indentation (compare Figure \ref{stress} e. with Figure \ref{stress} f.) this is due to the same normal force being distributed over a smaller contact area as only the leading edge of the indenter is in contact with material during sliding. Large zones of shear stresses are formed ahead of the contact under load (Figure \ref{stress} b. and d). These stress fields ahead of the contact explain the increase in hardness during scratching measured in \cite{kareer2016existence} and the interfacial slip phenomena observed at the edge of the contact \cite{brazil2021direct}.

When unloaded, this large shear stress disappears (compare positive shear right of the indenter in Figure \ref{stress}d. with Figure \ref{stress}i.) However, due to the plastic deformation and pile up of material at the edges and ahead of the scratch, significant residual stresses remain in the material once the load has been removed (compare Figure \ref{stress} b. with Figure \ref{stress} g.). Inside the scratch track, a residual shear stress of opposite sign to that ahead of the contact, forms in the wake of the contact due to the residual plastic deformation (negative shear stress in the wake of the contact of Figure \ref{stress} d.). As the scratch continues, a band of positive shear stress appears directly beneath this, once the material has elastically recovered. This band of positive shear stress remains once the load is fully removed and is validated by the residual HR-EBSD measurement shown in Figure \ref{stress} j.  The inhmogenous stress distribution aligns well with the hypothesis used to describe the formation of a dislocation trace line in \cite{Greiner2018TheOO}, where a parallel line of dislocations were observed just below the sliding surface. The shear stress ahead of the contact causes dislocations to move and are pushed into the surface (positive shear to the right of contact in Figure \ref{stress} d.), once the contact has passed, the opposite sign of the shear stress in the wake of the contact (negative shear to the left of the contact in (Figure \ref{stress} d.) causes dislocations to be pulled back up towards the surface, where they arrest in a self organised line at the region of zero shear stress between the positive and negative bands.

From this work, we can conclude that the stress field from tribological contact causes substantial plastic deformation, leading to material redistribution and the formation of material pile-up at the edges and ahead of the scratch track. These edge pile-ups do not occur in the absence of friction (i.e., during indentation alone), which can explain the observed inner and outer lattice rotation fields from our previous work that are not present in indentation where material piles up at the surface and the lattice rotates in one direction towards the indenter apex \cite{kareer2020scratching}. The dominant outer rotation field observed for a scratch, where the lattice rotates away from the indenter apex, accommodates this sliding-induced edge pile-up and the bulk of material removed from the scratch track is redistributed to the edges. 

The subsurface distribution of residual stresses under tribological loading is comparable to the elastic analytical solution in \cite{hamilton1983explicit} with a large shear stress ahead of the contact and a shear stress with opposite sign in its wake. However, the plastic deformation captured in this work results in residual shear stresses that consist of two bands of opposing signs and therefore a large gradient in the stress distribution below the surface. This stress gradient is likely to be a key mechanistic driver to the formation of the tribolayer and further microstructural evolution. In typical wear scenarios, the goal is to quantify the amount of material 'removed' from the surface whereas in the case of copper presented here, material is redistributed. However, by understanding the complex stress distribution induced during sliding, is it possible to identify points of strain localisation, from which it may be possible to predict the location and contact load at which a critical transition from plasticity to material detachment will occur. This fundamental understanding provides a pioneering framework from which multi-physical mechanisms may be integrated to accurately predict the wear response of a range of material system so that wear-resistant materials and coatings may be optimised and tailored to suit a specific application.




\bibliographystyle{elsarticle-num} 
\bibliography{report}

\begin{thebibliography}{10}
\expandafter\ifx\csname url\endcsname\relax
  \def\url#1{\texttt{#1}}\fi
\expandafter\ifx\csname urlprefix\endcsname\relax\def\urlprefix{URL }\fi
\expandafter\ifx\csname href\endcsname\relax
  \def\href#1#2{#2} \def\path#1{#1}\fi

\bibitem{jost1976economic}
P.~Jost, Economic impact of tribology, Proc mechanical failures prevention group (1976) 117--139.

\bibitem{holmberg2017influence}
K.~Holmberg, A.~Erdemir, Influence of tribology on global energy consumption, costs and emissions, Friction 5 (2017) 263--284.

\bibitem{lim1987wear}
S.~Lim, M.~Ashby, J.~Brunton, Wear-rate transitions and their relationship to wear mechanisms, Acta metallurgica 35~(6) (1987) 1343--1348.

\bibitem{lim1987overview}
S.~C. Lim, M.~Ashby, Overview no. 55 wear-mechanism maps, Acta metallurgica 35~(1) (1987) 1--24.

\bibitem{bowden2001friction}
F.~P. Bowden, D.~Tabor, The friction and lubrication of solids, Vol.~1, Oxford university press, 2001.

\bibitem{rigney1979plastic}
D.~Rigney, J.~Hirth, Plastic deformation and sliding friction of metals, Wear 53~(2) (1979) 345--370.

\bibitem{Greiner2018TheOO}
C.~Greiner, Z.~Liu, R.~Schneider, L.~Pastewka, P.~Gumbsch, The origin of surface microstructure evolution in sliding friction, Scripta Materialia (2018).

\bibitem{greiner2016sequence}
C.~Greiner, Z.~Liu, L.~Strassberger, P.~Gumbsch, Sequence of stages in the microstructure evolution in copper under mild reciprocating tribological loading, ACS applied materials \& interfaces 8~(24) (2016) 15809--15819.

\bibitem{Xu2021OnTO}
Y.~Xu, F.~Ruebeling, D.~S. Balint, C.~Greiner, D.~Dini, On the origin of microstructural discontinuities in sliding contacts: A discrete dislocation plasticity analysis, International Journal of Plasticity 138 (2021) 102942.

\bibitem{brazil2021direct}
O.~Brazil, G.~M. Pharr, Direct observation of partial interface slip in micrometre-scale single asperity contacts, Tribology International 155 (2021) 106776.

\bibitem{kareer2020scratching}
A.~Kareer, E.~Tarleton, C.~Hardie, S.~V. Hainsworth, A.~J. Wilkinson, Scratching the surface: Elastic rotations beneath nanoscratch and nanoindentation tests, Acta Materialia 200 (2020) 116--126.

\bibitem{kareer2016existence}
A.~Kareer, X.~Hou, N.~M. Jennett, S.~V. Hainsworth, The existence of a lateral size effect and the relationship between indentation and scratch hardness in copper, Philosophical Magazine 96~(32-34) (2016) 3396--3413.

\bibitem{kareer2016interaction}
A.~Kareer, X.~Hou, N.~M. Jennett, S.~Hainsworth, The interaction between lateral size effect and grain size when scratching polycrystalline copper using a berkovich indenter, Philosophical Magazine 96~(32-34) (2016) 3414--3429.

\bibitem{hamilton1983explicit}
G.~Hamilton, Explicit equations for the stresses beneath a sliding spherical contact, Proceedings of the Institution of Mechanical Engineers, Part C: Journal of Mechanical Engineering Science 197~(1) (1983) 53--59.

\bibitem{wilkinson2006high}
A.~J. Wilkinson, G.~Meaden, D.~J. Dingley, High-resolution elastic strain measurement from electron backscatter diffraction patterns: New levels of sensitivity, Ultramicroscopy 106~(4-5) (2006) 307--313.

\bibitem{wilkinson2010determination}
A.~J. Wilkinson, D.~Randman, Determination of elastic strain fields and geometrically necessary dislocation distributions near nanoindents using electron back scatter diffraction, Philosophical magazine 90~(9) (2010) 1159--1177.

\bibitem{britton2012high}
T.~Britton, A.~J. Wilkinson, High resolution electron backscatter diffraction measurements of elastic strain variations in the presence of larger lattice rotations, Ultramicroscopy 114 (2012) 82--95.

\bibitem{dunne2007lengthscale}
F.~Dunne, D.~Rugg, A.~Walker, Lengthscale-dependent, elastically anisotropic, physically-based hcp crystal plasticity: Application to cold-dwell fatigue in ti alloys, International Journal of Plasticity 23~(6) (2007) 1061--1083.

\bibitem{dunne2012crystal}
F.~Dunne, R.~Kiwanuka, A.~Wilkinson, Crystal plasticity analysis of micro-deformation, lattice rotation and geometrically necessary dislocation density, Proceedings of the Royal Society A: Mathematical, Physical and Engineering Sciences 468~(2145) (2012) 2509--2531.

\end{thebibliography}







\end{document}